\documentclass[10pt,twocolumn,letterpaper]{article}

\usepackage{cvpr}              

\usepackage{graphicx}
\usepackage{amsmath}
\usepackage{amssymb}
\usepackage{booktabs}

\usepackage[pagebackref,breaklinks,colorlinks]{hyperref}

\usepackage[capitalize]{cleveref}
\crefname{section}{Sec.}{Secs.}
\Crefname{section}{Section}{Sections}
\Crefname{table}{Table}{Tables}
\crefname{table}{Tab.}{Tabs.}

\def\cvprPaperID{*****} 
\def\confName{CVPR}
\def\confYear{2022}

\begin{document}

\title{Hyper-Convolutions via Implicit Kernels for Medical Imaging}

\author{Tianyu Ma\\
Cornell University\\
Cornell Tech\\ 
{\tt\small tm478@cornell.edu}
\and 
Alan Q. Wang \\
Cornell University\\
Cornell Tech\\ 
{\tt\small aw847@cornell.edu}
\and 
Adrian V. Dalca\\
Massachusetts Institute of Technology\\
Massachusetts General Hospital\\
Harvard Medical School\\
{\tt\small adalca@mit.edu}
\and 
Mert R. Sabuncu\\
Cornell University\\
Cornell Tech\\ 
{\tt\small msabuncu@cornell.edu}}
\maketitle

\begin{abstract}
The convolutional neural network (CNN) is one of the most commonly used architectures for computer vision tasks. 
The key building block of a CNN is the convolutional kernel that aggregates information from the pixel neighborhood and shares weights across all pixels. 
A standard CNN's capacity, and thus its performance, is directly related to the number of learnable kernel weights, which is determined by the number of channels and the kernel size (support). 
In this paper, we present the \textit{hyper-convolution}, a novel building block that implicitly encodes the convolutional kernel using spatial coordinates. 
Hyper-convolutions decouple kernel size from the total number of learnable parameters, enabling a more flexible architecture design. 
We demonstrate in our experiments that replacing regular convolutions with hyper-convolutions can improve performance with less parameters, and increase robustness against noise.
We provide our code here: \emph{https://github.com/tym002/Hyper-Convolution}
\end{abstract}

\section{Introduction}
\label{sec:introduction}
Convolutional neural networks (CNNs) are widely used in computer vision and medical imaging.
The core ingredient of a CNN is the convolutional kernel, which has a fixed number of learnable weights that is proportional to the kernel size. 
For many computer vision tasks, especially in the biomedical domain, long-range dependencies can be critical~\cite{parsnet,Samy2018nunet}. 
In CNNs, successive convolutional layers, increased kernel size, and downsampling operations are used to capture long-range information~\cite{Krizhevsky2012alex,Lecun1989backprop,unet,Simonyan2014vgg}.

Larger convolution kernels have more learnable parameters, which can result in overfitting when training data are limited, as is the case in many biomedical applications.
Alternative kernel representations, such as deformable~\cite{dai2017deformable,thomas2019kpconv} and dilated convolutions ~\cite{dilated3, dilated2,dilated1} can mitigate this issue. 
These methods, however, are often more effective in tasks such as image classification and detection. 
For tasks like image segmentation and reconstruction, which require dense pixel-level predictions, sparse kernels are often insufficient \cite{Huang_2019_CCnet}.

\begin{figure}[!t]
\centerline{\includegraphics[width=0.8\linewidth]{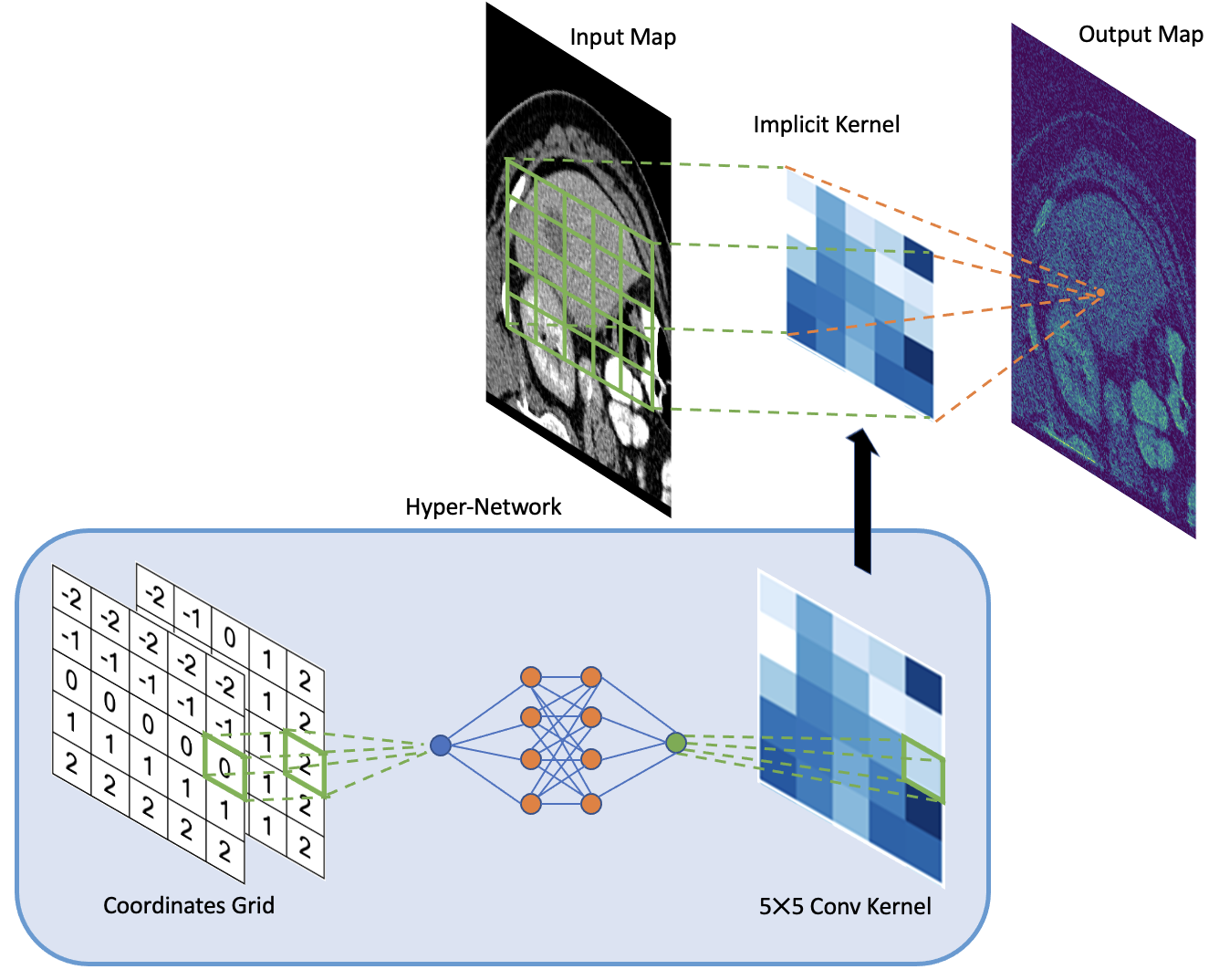}}
\caption{Illustration of hyper-convolution and the implicit kernel. A hyper-network takes kernel coordinates and produces kernel weights. The only learnable weights are in the hyper-network, independent of the size of the used kernel.}
\label{fig1}
\end{figure}

Another limitation of regular convolution is the lack of robustness~\cite{schneider2020improving}. 
A CNN will often achieve good performance when train and test distributions are close. 
However, these models can be highly susceptible to perturbations on test images, for example due to noise, compression artifacts, and adversarial attacks \cite{szegedy2013intriguing}. 
In biomedical imaging techniques such as MRI, scanner noise and distributional shift is a fairly common challenge~\cite{gudbjartsson1995rician}.

In this paper, we present a novel building block for CNNs that we call \textit{hyper-convolution}.
Hyper-convolution is an implicit representation of a convolutional kernel as a parametrized function of kernel spatial coordinates. Instead of learning all kernel weights independently during training, a hyper-convolution learns a functional mapping that computes weights from grid coordinates. This method decouples the number of learnable parameters from the size of the kernel.   
We illustrate the proposed hyper-convolution building block in Figure~\ref{fig1}.
Similar to a regular convolution, the hyper-convolution implements a dense kernel, yet we can modify its expressiveness and receptive field without changing the number of learnable parameters.
This affords us additional flexibility in designing convolutional architectures. 
We demonstrate in our experiments that, in contrast to regular convolutional kernels, implicit kernels tend to be spatially smooth due to its implicit parameterization and altered learning dynamics. 
As a result, models with hyper-convolution can be more robust. 
We perform experiments on both image segmentation and reconstruction tasks: with two challenging biomedical image segmentation datasets and one brain MRI reconstruction dataset.  All experiments show improved performance and robustness against noise during testing.
This work is a substantial extension of our prior conference paper~\cite{ma2022hyper}. 
Here, we present more theoretical details, novel experimental results, and expanded Discussion and Background sections.

\section{Related Works}

\subsection{Alternative Convolutional Kernels}
The impact of large convolutional kernels have been studied for image segmentation, classification and object detection \cite{cai2018proxylessnas, large, tan2019mixconv, tan2019mnasnet}.
Symmetric and separable convolutional filters have been employed to reduce computational cost and number of parameters \cite{chollet2017xception, martucci1994symmetric}. 
Dilated convolutions are popular for increasing the receptive field of a neural network without increasing the number of learnable parameters ~\cite{yu2015dilation}.
Adapting the dilated convolution concept, atrous spatial pyramid pooling was developed to aggregate long-range dependencies in images at multiple scales, and yielded good performance on several image segmentation datasets \cite{deeplab,dilated3,dilated2,unetdilated}. 
Another popular technique to increase model capacity is the deformable convolution \cite{dai2017deformable,zhu2019deformable}, which learns the sampling locations of kernels and adapts to geometric variations in objects.  

Although the number of learnable parameters can be reduced by these alternatives, they maintain a strong coupling between model capacity and the number of pixels used in the kernel. 
This coupling can lead to overfitting or limit the expressiveness for a wide variety of tasks that require capturing long-range dependencies and producing dense, pixel-level predictions.  
In contrast, the proposed hyper-convolution breaks the link between kernel size and the learnable weights, thus enabling the flexibility to design more effective networks. 

\subsection{Non-local Networks}

Recently, inspired by the success of transformers in natural language processing~\cite{vaswani2017attention}, self-attention and non-local networks have gained popularity in computer vision, in part, due to their ability to aggregate long-range dependencies~\cite{dosovitskiy2020image, liu2021swin, wang2021pyramid}. 
A non-local block allows the receptive field of the network to be the entire image \cite{vaswani2017attention,nonlocal}.
However, this can be computationally very expensive.
Some works have focused on improving the efficiency of non-local layers, such that they can be used with high resolution image grids~\cite{gnonlocal,asymmetric}. 
Non-local operations have also been used in CNNs, for example, at the bottleneck of the UNet architecture~\cite{nonlocalunet}.
%
In contrast to self-attention layers, the proposed hyper-convolution technique is feasible at any resolution without a heavy computation or memory overhead. 

\subsection{Hyper-Networks and Implicit Repreentations}
Hyper-networks generate weights for another network responsible for the main task, and have been used in a wide variety of fields including computer vision and natural language processing~\cite{suarez2017language}. 
Hyper-networks are powerful tools that can improve parameter-efficiency and afford flexibility, without significantly sacrificing expressiveness~\cite{hypernet}. 
For example, the HyperSeg \cite{nirkin2021hyperseg} architecture encodes the input image by a hyper-network, which in turn generates the weights of a decoder that solves a segmentation task.
Hyper-networks have also been used to obtain models that are agnostic to the degree of regularization for image registration and reconstruction tasks~\cite{hyperpara,recon}. 

Neural networks have also been used to create an \textit{implicit} representation of signals, such as natural images, for instance, to achieve compression. 
For example, a small neural network can be trained to map pixel coordinates to RGB-values, where an image is represented in the learned weight values~\cite{represent,Sitzmann2020siren}.
This approach can also be used to learn representations of 3D shapes~\cite{sal,shape}.
More similar to our work, implicit representations can be employed to define kernel functions for irregularly structured point cloud data~\cite{wang2018cloud}. 

\subsection{Robust Convolutional Neural Networks}
CNNs are particularly successful when training and testing data are drawn from the same distribution without significant domain shift~\cite{sun2015survey}. 
However, CNNs are also known to be non-robust when there are perturbations in the test images, even when such changes are very small. 
To address this, Schneider et el. \cite{schneider2020improving} proposed using covariate shift adaptation.
Noise injection and data augmentation during training are also common strategies to improve robustness~\cite{rusak2020simple,yin2019fourier}. 
Relevant to our work,  the use of smooth convolutional kernels have been shown to be more robust against noise and adversarial attacks~\cite{smooth,wang2020high}.

\section{Proposed Method}
The hyper-convolution is a building block that can replace a regular convolution in a CNN. 
The core idea of hyper-convolution is to generate all the weights in a convolutional kernel using a neural network, which can be seen as an implicit representation for the kernel. The input to the neural network is the kernel's spatial coordinates. 

For a standard convolutional kernel, all the trainable weights are independent and explicitly learned during training. 
Instead, in hyper-convolutions, the values of all the weights in a kernel are implicitly coupled via the functional representation of the hyper-network. 
Unlike the standard convolution operation, the size of the convolutional kernel is a design choice that does not affect the number of learnable parameters.

Specifically, a hyper-convolution is a function $\Phi_\theta(\cdot)$, parameterized by $\theta$, that maps kernel grid coordinates to filter weights. For example, for a 2D convolutional kernel, 
\begin{equation}
K_{ij} = \Phi_\theta(i,j)
\end{equation}
where $(i,j) \in \mathbb{R}^2$ and $K_{ij}$ indicates the filter weight at filter location $i, j$. In our implementation, the center pixel of the convolution kernel is the origin and has coordinates $(0,0)$. 

\subsection{Hyper-network Architecture}


For each layer in the main task CNN\footnote{Except for the final output, which is a $1\times1$ convolutional layer.}, we utilize a multi-layer fully-connected hyper-network which takes coordinates as input. 
For computational efficiency, we implement the hyper-network as a CNN $\Phi_\theta(\cdot)$ with $1\times 1$ convolution layers and leaky ReLU nonlinearities of slope 0.1~\cite{maas2013rectifier}.
In our implementation, the input is a coordinate grid with the same size as the convolutional kernel in the main network.
Because all the trainable parameters are in the hyper-network instead of the convolutional kernels, the capacity of $\Phi_\theta$ determines the expressiveness of the hyper-convolutional kernels. 
If the number of parameters in the hyper-network is significantly smaller than the kernel size, the model is more restricted.   
In all of our experiments, the hyper-network we use has four hidden layers before outputting the final weights. The first three layers have a fixed number of nodes, and the number of nodes $N_L$ for the final hidden layer is a hyper-parameter that dictates the capacity of the hyper-networks. 

\subsection{Parameter Efficiency of Hyper-convolution}
One of the advantages of hyper-convolution is the decoupling of the kernel size and the number of trainable parameters. 
In a standard convolution layer, where the 2D kernel size is $h\times w$ (e.g., $3 \times 3$), and the numbers of input and output channels are $N_{in}$ and $N_{out}$, the total number of parameters is:
\begin{equation}
(h \times w) \times N_{in} \times N_{out},
\end{equation}
excluding the bias terms. The total number of parameters increases quadratically with the size of the convolutional kernel.  
For hyper-convolution, the total number of parameters with $L$ layers in $\Phi_\theta$ can be calculated as: 
\begin{equation}
(N_L+1) N_{in} N_{out}  +  \sum_{j=0}^{L-1} (N_j+1)N_{j+1}
\end{equation}
where $N_j$ is the number of nodes in the $j$'th layer, and $N_L$ is the number of channels of the final hidden layer in the hyper-network. 
The number of learnable parameters of the hyper-convolution block is independent of the kernel size $h\times w$, and depends on the number of input and output channels in the main network, as well as the hyper-parameter $N_L$. 

In practice, for most convolutional neural networks the number of channels (particularly in the hidden layers) $N_{in}$ and $N_{out}$ are often chosen to be $8$ or larger. 
Furthermore, in our hyper-network design, we choose the number of nodes $N_j$ before the penultimate layer $N_L$ to be small (e.g. less than 8). 
Under these conditions, the number of parameters in the proposed hyper-convolution layer is dominated by the final hidden layer of the hyper-network and is approximately:
\begin{equation}
(N_L+1) \times N_{in} \times N_{out}.
\end{equation}

If $N_L < h\times w$, the hyper-convolution has fewer parameters than a standard convolution kernel. 
This way, for a fixed number of parameters, the proposed method can implement dense kernels with arbitrarily large receptive fields, capturing high-resolution contextual information in features. 
Hyper-convolutions can thus afford an expanded receptive field with high capacity, without increasing the number of trainable parameters, thus limiting the risk of overfitting. 
Because the input to the hyper-network is a small coordinate grid, e.g. $5\times5$, the additional flops and memory required for hyper-convolution is much smaller than those of the main network. 
Thus, the majority of the memory and computational burden is due to the main network. 
For example, in our experiments, inference time for a UNet with $5 \times 5$ kernels and a corresponding hyper-UNet with same kernel size are $14$~ms and $16$~ms per 2D image, respectively.

\subsection{Robustness of Hyper-Convolutions} \label{robust}
In a regular convolution layer, each kernel weight is typically randomly initialized and updated independently. 
In contrast, in the hyper-convolution, due to the implicit representation, there is a correlation between kernel weight values, both at initialization and during gradient-based updates.
This correlation thus regularizes the kernel weights, making them spatially more smooth, which, in turn, can play an important role in robustness.
We provide an exploration of this hypothesis in our experiments.  

\section{Experiments}
\begin{figure}
\begin{center}
\includegraphics[width=0.9\linewidth]{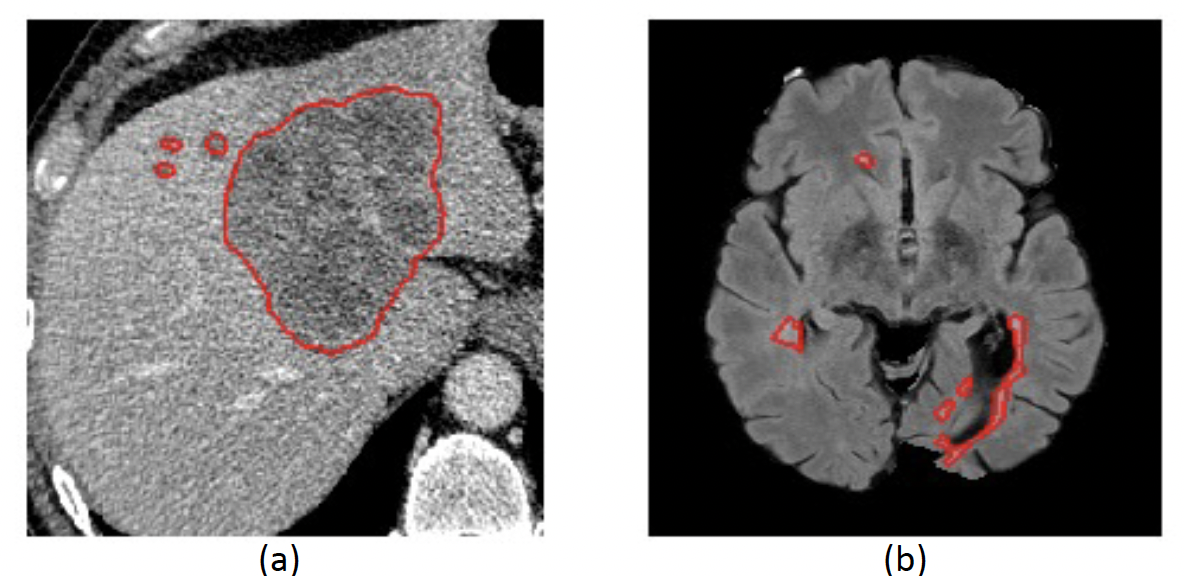}
\vspace{-0.5cm} 
\end{center}
   \caption{Example images for segmentation tasks we consider. (a): Liver lesion. (b): MS-lesion.}
\label{fig:exp}
\end{figure}

In our experiments, we focus on biomedical imaging applications and demonstrate the effectiveness of our method for segmentation and image reconstruction.
We choose these tasks because they can benefit from an increased receptive field, yet are prone to overfitting due to limited training examples.
For image segmentation, we consider two problems: liver lesion segmentation \cite{lits} and Multiple Sclerosis (MS) lesion segmentation \cite{mslesion} (see Figure~\ref{fig:exp}).
For reconstruction, we use T1-weighted axial brain MRI images from the ABIDE dataset~\cite{abide}. 

For all experiments, we perform data augmentation, including vertical and horizontal flipping, random rotation up to $30$ degrees, and scaling between $0.9$ and $1.1$.  
Adam optimizer \cite{adam} with a learning rate of 0.0001 and a mini-batch size of 8 (for liver lesion segmentation and brain MRI reconstruction) or 16 (for MS lesion segmentation) are used for training. 
We also use dropout regularization with 0.5 probability and batch normalization. 

For segmentation we use soft Dice loss \cite{milletari2016vnet} for training and report Dice score results for the epoch with the best validation loss. 
The Dice score quantifies the overlap between the automatic and ground-truth manual segmentations, and is widely used in the literature \cite{ma2021ensembling}.
We use mean-square error (MSE) as the loss function for image reconstruction and also report peak signal-to-noise ratio (PSNR) values.

\subsection{Baselines} 
As noted above, implicit kernels can be used to replace regular convolutions in any CNN architecture. 
In our experiments, we explore different CNN architectures as baselines. 
The UNet backbone, a popular architecture, is used for both segmentation and reconstruction tasks. 
The UNet adopts a multi-scale feature learning, aggregating contextual information using multiple max-pooling operations as well as skip connections between the down- and up-sampling paths~\cite{unet}. 
Our 2D UNet has three max-pool layers with two convolution layers per scale, and ReLU nonlinearities, with the number of channels doubled after each max-pool layer. 
Below, when we talk about the number of channels of a UNet model, we refer to the number of channels of the first convolutional layer. 
To gain a more comprehensive understanding of the segmentation performance, we experiment with different variations of the UNet backbone. 
We vary the kernel size ($3\times 3$ and $5\times 5$) used in the UNet and replace regular kernels with dilated convolutions~\cite{unetdilated}. 
We also experiment with modifying the number of channels, as described in the results sections.
For competing methods, we implement a non-local UNet \cite{nonlocalunet}, which integrates a non-local self-attention block into the bottleneck. The non-local UNet also serves as one of our backbone methods as we replace convolutions in the architecture with hyper-convolutions.
Additionally, we include an implementation of HyperSeg\cite{nirkin2021hyperseg}, a recent segmentation method employing hyper-networks.

For the segmentation task, we also implement a flat CNN architecture. 
The flat CNN backbone \cite{yu2015dilation} uses a series of convolutional layers with different kernel sizes that gradually expand and then shrink: $3\times3$, $5\times5$,  $9\times9$, $5\times5$, and $3\times3$.
The number of channels for each convolutional layer is equal to the kernel size. E.g., for the layer with a $9\times9$ kernel, we have 81 output channels.
Different from the UNet and other popular convolutional neural networks, this architecture does not have any down/up-sampling layers and all convolution operations are executed at original resolution, followed by batch-normalization and activation. 
As another baseline, we implement a 2D flat CNN with dilated convolutions that gradually expand the receptive field consisting of sequential residual $3\times3$ convolutional kernels, with dilations of 1, 2, 4, 8, 4, 2, 1. 
The numbers of channels for each of these convolutional layers are 16, 32, 64, 128, 64, 32, 16, respectively.

\subsection{Segmentation with Clean Image}
We first run experiments to examine the performance of hyper-convolutions with noise-free input images. We experiment with noisy images in Section~\ref{seg_noisy}.

\subsubsection{Liver lesion segmentation}
\textbf{Data:}
We use the LiTS dataset \cite{lits} for the liver lesion segmentation task, which includes 131 liver Computed Tomography (CT) volumes with ground truth manual segmentations.
We randomly split the data: 80 cases for training, 20 for validation, and 31 for testing.
For each 3D CT volumes, the number of 2D slices (of size $512\times 512$) varies between 74 and 987. There are in total 58638 2D slices for training, validation and testing. 
We pre-process all slices by resizing them to $256\times 256$ and truncate the intensity range to $[-100,250]$ before finally mapping it to $[0,1]$. 

\begin{figure}[t]
\begin{center}
\includegraphics[width=1\linewidth]{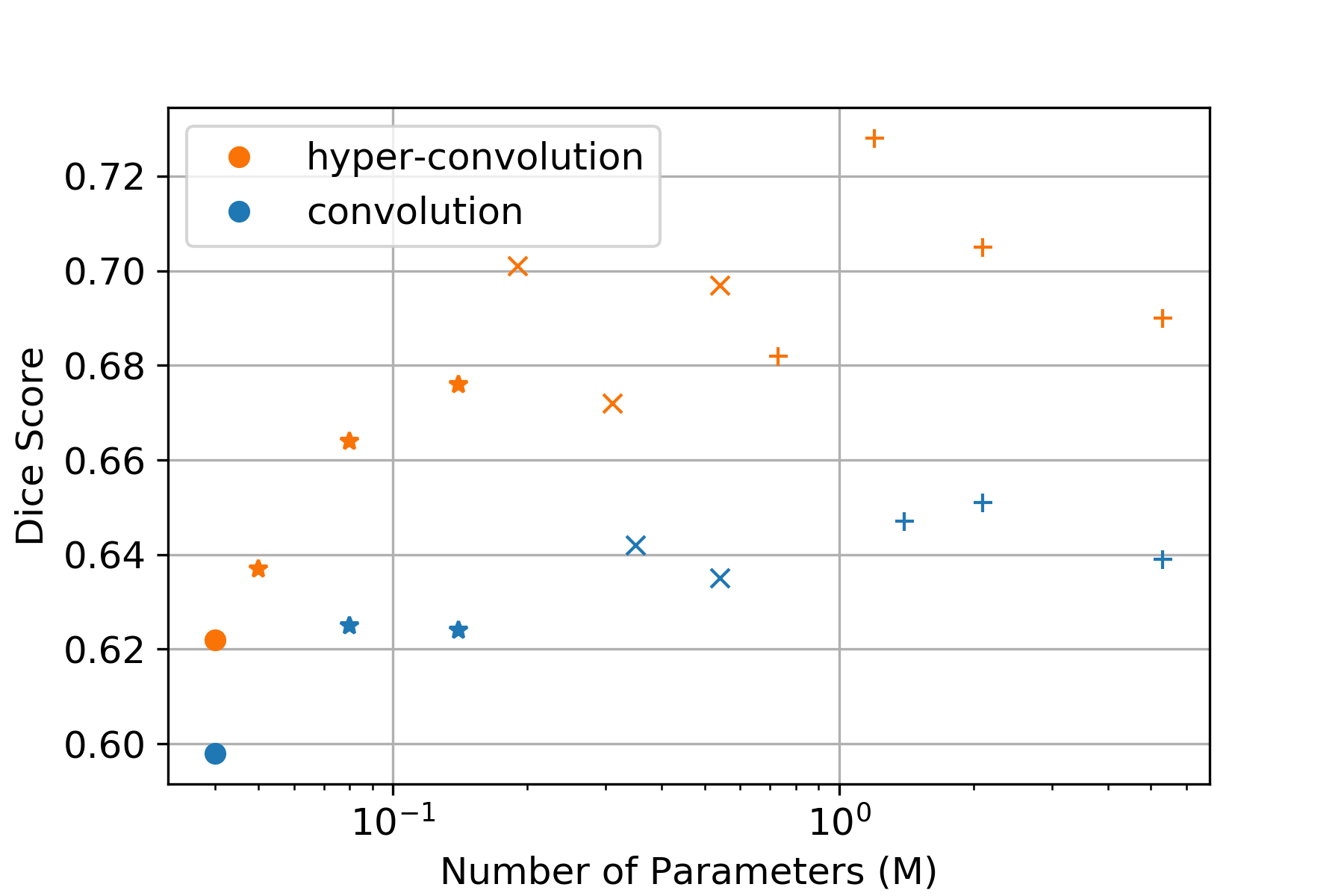}
\end{center}
   \caption{Liver Segmentation Test Dice scores:
   $5\times 5$ Hyper-UNet  (blue) and standard UNet (red)  with different numbers of parameters in millions. $\circ$,$\star$,$\times$, and + indicates 4,8,16 and 32 initial channels.}
\label{fig:liver}
\end{figure}

\begin{table*}
\begin{center}
\begin{tabular}{|l|c|c|c|c|}
\hline
Method &  Train Dice & Test Dice & Receptive Field & Params (M)\\
\hline\hline
UNet $3\times 3$ \cite{unet} &  0.931 & 0.651 & 68 pixels & 2.1\\
UNet $5\times 5$ &  0.942 & 0.639 & 128 pixels & 5.3\\
Dilated UNet $3\times 3$ \cite{unetdilated} & 0.930 & 0.612 & 128 pixels & 2.1\\
HyperSeg \cite{nirkin2021hyperseg} & 0.902 & 0.683 & All pixels & 1.2 \\
Non-local UNet $3\times 3$ \cite{nonlocalunet} & 0.919 & 0.690 & All pixels & 2.3\\
Hyper-UNet $5\times 5$ (ours) & 0.886 & $0.728$ & 128 pixels & 1.2\\
Non-local Hyper-UNet $5\times 5$ (ours) & 0.893 & $\mathbf{0.733}$ & All pixels & 1.4\\
\hline
Flat Dilated CNN \cite{yu2015dilation} & 0.892 & 0.607 & 89 pixels & 0.45\\
Flat Hyper-CNN (ours) & 0.824 & 0.647 & 89 pixels & 0.45\\
\hline
\end{tabular}
\end{center}
\caption{Train and test performance of different models in the liver lesion segmentation task. Best test Dice score is \textbf{bold-faced}.}\label{tab1}
\vspace{-0.5cm}
\end{table*}

\textbf{Results:}
The first set of experiments are done by replacing all convolution kernels in a UNet with hyper-convolutions of the same size ($5\times5$). 
We vary the number of channels to compare their performances across a wide range of number of parameters. 
In addition to the number of channels in the main network, we train different hyper-UNet models by varying the number of units  in the penultimate layer of the hyper-network, $N_L$. The Dice scores achieved by the UNet and Hyper-UNet models are shown in Figure~\ref{fig:liver}.   

We observe that hyper-convolutions yield a consistent and significant performance boost across a wide range of total number of learnable parameters. For both UNet and hyper-UNet, the performance increases with the number of parameters due to higher capacity models. 
However, we note that increasing the number of parameters beyond some threshold can yield a drop in performance, likely because of  overfitting.  

We also compare the proposed method with other architectures and show all the results in Table~\ref{tab1}, which lists 
 training and testing results obtained at the epoch with best validation loss, in addition to the size of receptive field and the number of learnable parameters, for different baseline models and their hyper-convolution counterparts.
We first observe that convolutional architectures can benefit from expanded receptive fields that use larger kernels. 
As we increase the kernel size from 3 to 5 in the UNet baseline, the train Dice improves, consistent with increased model capacity.
However, simply expanding the kernel size from 3 to 5 makes the total number of learnable parameters more than double. The increasing difference between the test and train Dice scores is direct evidence for more severe overfitting due to the increase in the number of parameters.
We also include a dilated UNet baseline with $3\times3$ kernels and a dilation of 2, which has the same receptive field as a $5\times5$ UNet. 
Despite having the same number of learnable parameters as a $3\times 3$ UNet, the dilated UNet has worse test performance, indicating that simply increasing the receptive field does not necessarily improve model performance for this task.  
By utilizing a self-attention mechanism, the non-local UNet baseline shows a robust improvement in test Dice score, without a significant increase in the total number of parameters. 
This performance boost is likely due to the non-local UNet's capability of utilizing both global and local information to make predictions. 

The Hyper-UNet and non-local Hyper-UNet (included in Table~\ref{tab1}) modify the UNet and non-local UNet backbones by replacing all convolutions with $5 \times 5$ hyper-convolutions with $N_L=4$. 
These hyper-convolutions provide an increased receptive field with only half of the total number of learnable parameters as the $3\times3$ UNet baseline. 
The receptive field of hyper-UNet is doubled (128 pixels) compared to the UNet baseline, covering one half of the entire image. 
In addition to the larger field-of-view, hyper-convolutions have a more restricted kernel, which effectively acts as regularization.
We observe that Hyper-UNet and Non-local Hyper-UNet exhibit less overfitting and achieve the best test performance.

The flat CNN baseline with dilated convolutions has only 0.45M learnable parameters - much less than the UNet counterparts, but a larger receptive field of 89 pixels.   
Its test performance, however, is the worst among all baselines, due to substantial overfitting. 
In contrast, the flat Hyper-CNN, which has the same number of learnable parameters and the same receptive field size as the flat baseline, achieves a test Dice score that is comparable to the $3\times3$ UNet. 
We note therefore that this performance improvement is in part due to the regularization effect of hyper-convolutions and is not merely due to the number of parameters.


\begin{table}
\begin{center}
\begin{tabular}{|c|c|c|c|}
\hline
Size, $N_L$  & Test Dice & Recep. Field & Params (M)\\
\hline\hline
$3\times 3$, $2$  & 0.604 & 68 pixels & 0.73\\
$3\times 3$, $4$  & 0.627 & 68 pixels & 1.2\\
$3\times 3$, $8$  & 0.648 & 68 pixels & 2.2\\
\hline
$5\times 5$, $2$  & 0.692 & 128 pixels & 0.73\\
$5\times 5$, $4$  & 0.728 & 128 pixels & 1.2\\
$5\times 5$, $8$  & 0.705 & 128 pixels & 2.2\\
\hline
$7\times 7$, $2$  & 0.683 & 188 pixels & 0.73\\
$7\times 7$, $4$  & 0.704 & 188 pixels & 1.2\\
$7\times 7$, $8$  & 0.717 & 188 pixels & 2.2\\

\hline
\end{tabular}
\end{center}
\caption{Performance of Hyper-UNet on Liver Lesion data with different kernel sizes and hyper-network capacity.}\label{tab2}
\vspace{-0.5cm}
\end{table}

\begin{figure}[t]
\begin{center}
\includegraphics[width=0.7\linewidth]{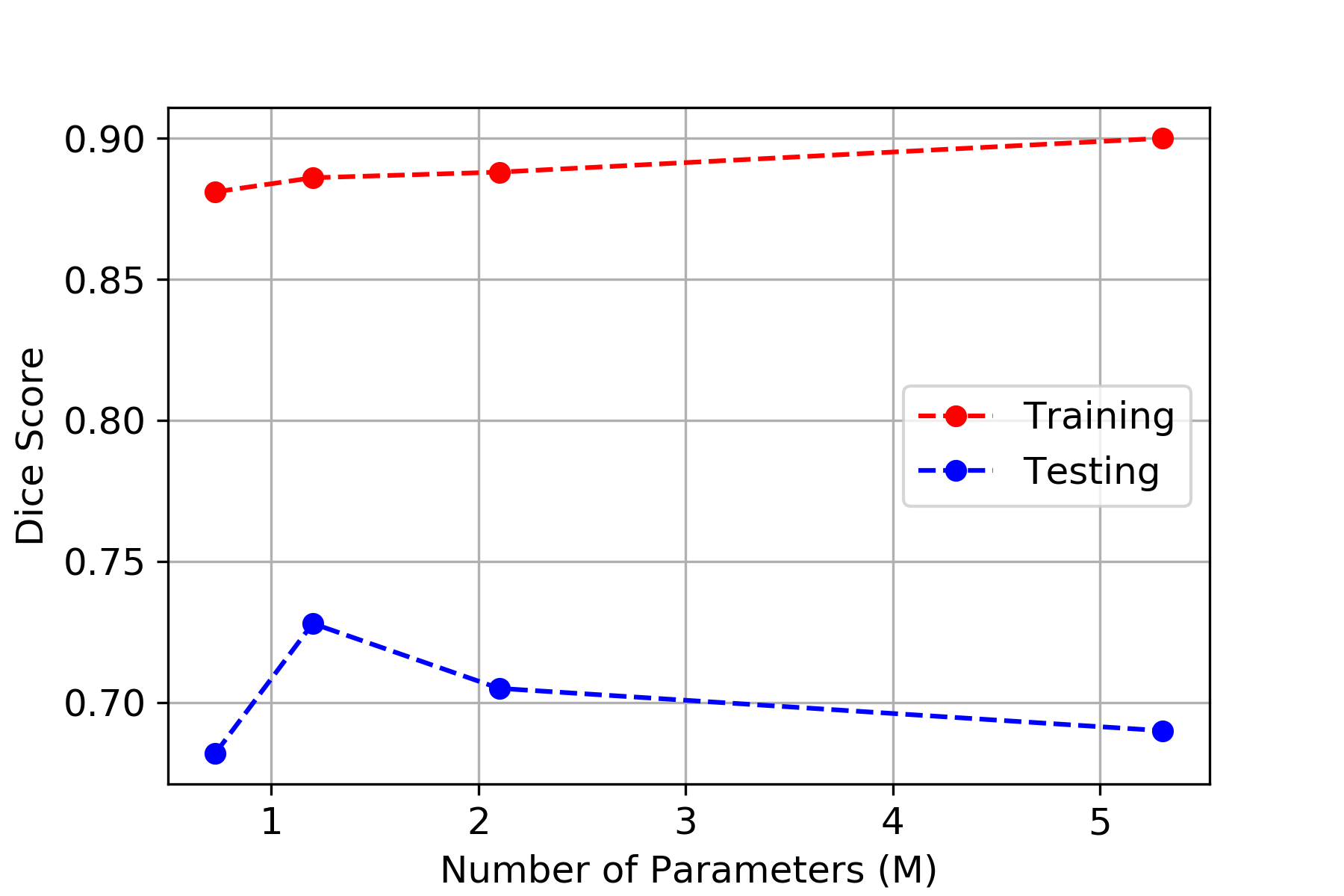}
\vspace{-0.5cm}
\end{center}
   \caption{Train (blue) and test (orange) Dice scores for $5\times 5$ Hyper-UNet with different numbers of parameters in millions, corresponding to $N_L = 2, 4, 8,  24$.
   \vspace{-0.5cm}}
\label{fig:traintest}
\end{figure}

\begin{figure}[tb]
\begin{center}
\includegraphics[width=0.7\linewidth]{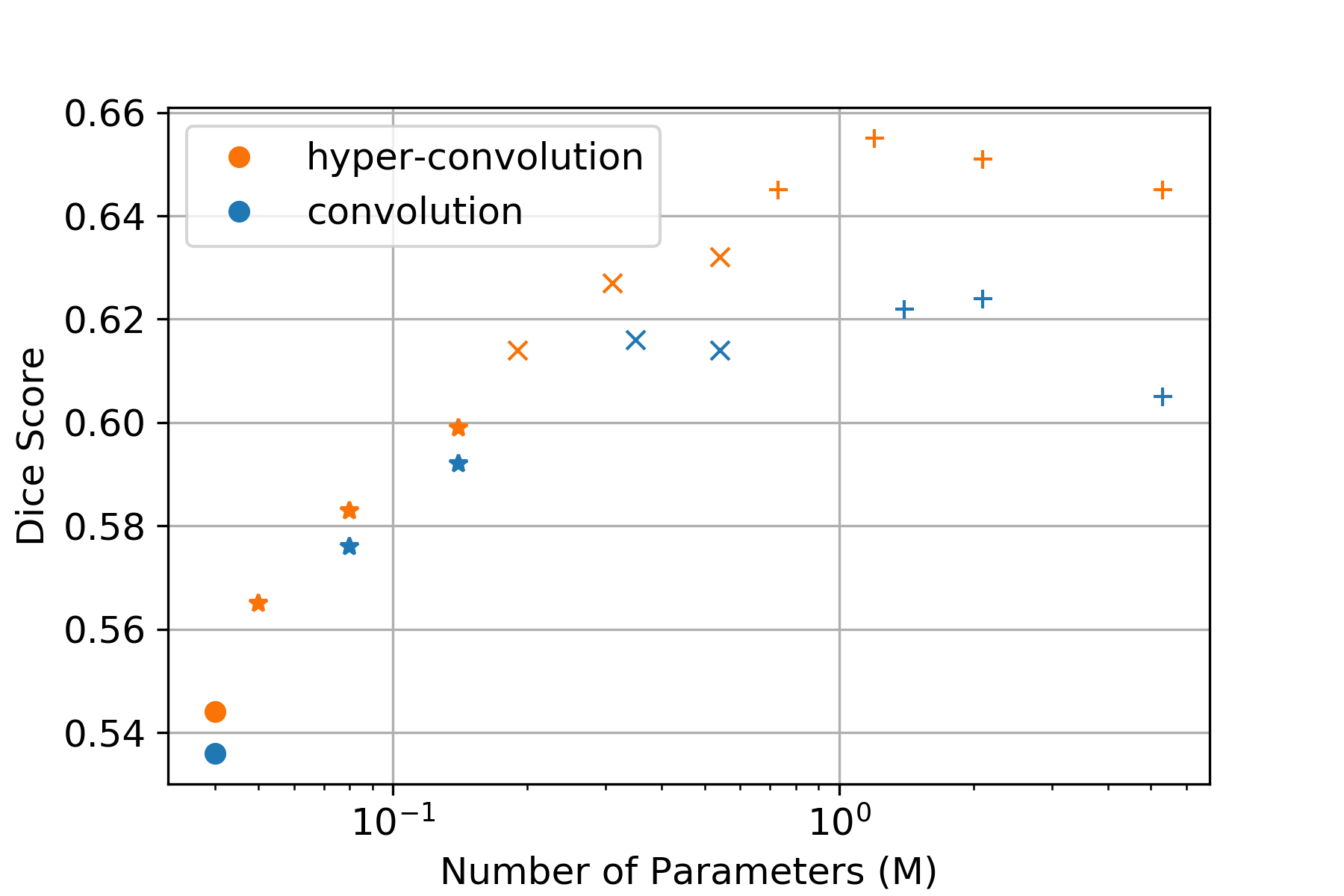}
\end{center}
   \caption{Test Dice scores for $5\times 5$ Hyper-UNet (blue) and UNet baseline (red) with different numbers of parameters. $\circ$,$\star$,$\times$, and + correspond to 4,8,16 and 32 initial channels in the backbone architecture.}
\label{brain}
\end{figure}




\begin{figure*}[tbh]
\begin{center}
\includegraphics[width=0.9\linewidth]{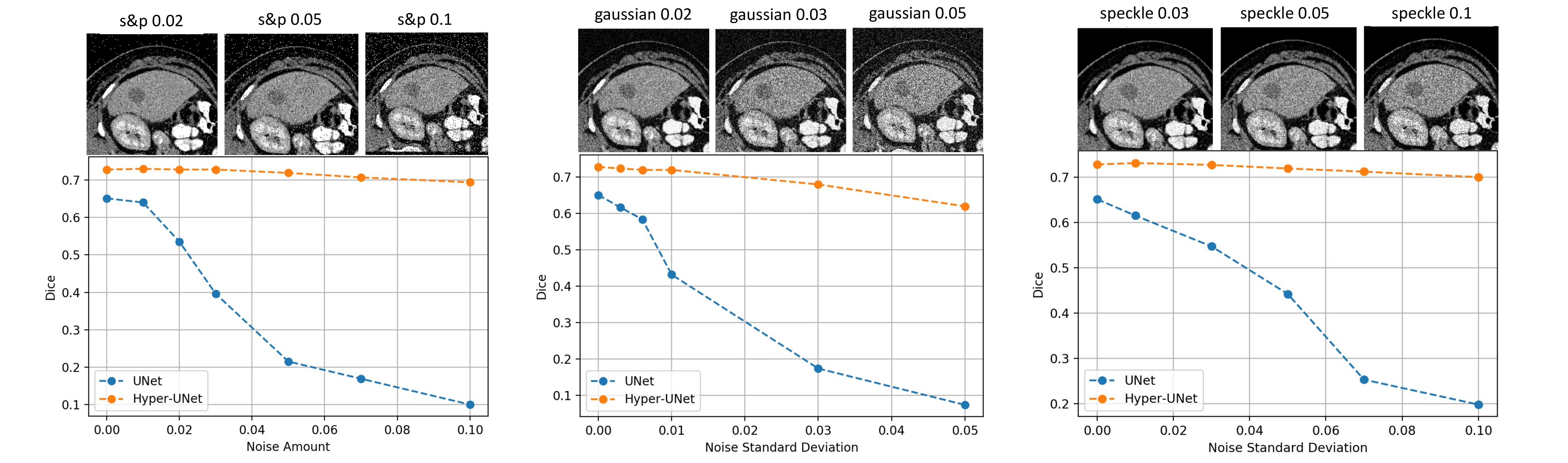}
\end{center}
   \caption{Top: Example test images with salt and pepper noise, Gaussian noise, and Speckle Noise. Bottom: Liver segmentation average test Dice scores for different noise levels.}
\label{noise}
\end{figure*}

\textbf{Varying Hyperparameters:} 
We run a set of experiments to further interrogate the effect of kernel size and $N_L$ on performance.
Recall that for hyper-convolution, the kernel size is decoupled from the total number of trainable parameters. 
Table~\ref{tab2} shows results for the 32-channel Hyper-UNet with variable kernel sizes and $N_L$ values. 
We observe that the $3\times 3$ implicit kernel performs worse than the standard $3\times 3$ convolution, particularly when the capacity is restricted with small $N_L$. 
The gap between hyper-convolution and regular convolution shrinks as we increase $N_L$ and almost disappears when they have the same number of parameters. 
Larger kernel sizes achieve better test Dice scores, however increasing hyper-network capacity (i.e., $N_L$) does not always improve performance (seen with $5\times5$ kernels).
The diminishing return of increasing kernel size is likely due to overfitting of the higher capacity model. 
Additionally, because the input image only has a size of 256 pixels, the receptive field of a $5\times5$ kernel might already be sufficient.  

Figure~\ref{fig:traintest} shows the train and test Dice scores for $5\times 5$ hyper-convolutions with different $N_L=2, 4, 8, 24$. 
A hyper-convolution model with $N_L=24$ has approximately the same number of learnable parameters as a regular $5 \times 5$ UNet. 
We note that the train Dice increases with number of parameters due to larger model capacity. 
However, test Dice peaks at $N_L=4$, demonstrating that regularization of hyper-convolution can improve generalization. 

\subsubsection{Multiple Sclerosis Lesion Segmentation}
\textbf{Data:}
Next, we consider a Multiple Sclerosis (MS) lesion segmentation task.
We use a public dataset~\cite{mslesion}, which contains brain MRI scans from 19 subjects, each with 4-6 scans from different time points. 
Among the 19 subjects, manual annotations are provided for 5 subjects (21 images), and the remaining 14 subjects (61 images) are used for independent testing.  
For the 14 test subjects, we do not have access to the ground truth segmentation, and we submit model outputs via an online portal for evaluation, which returns test Dice scores.
The test results reported below are compiled from these returned test Dice scores.
The dataset has two independent expert annotations, which have a Dice overlap score of 0.732, highlighting how challenging the task is. 
We use the intersection of the two manual labels as the gold standard labels during training and validation. 
Each image contains four different Magnetic Resonance Imaging (MRI) contrasts: FLAIR, PD-weighted, T2-weighted, and T1-weighted. 
The original images have size $182 \times 256 \times 182$, totaling 3822 2D images for training. 
Images were pre-processed by cropping the center of each 3D image to $144 \times 176 \times 144$ and applying z-score normalization.

Since we only have 5 subjects with gold standard segmentations available for training and validation, we run 5-fold experiments where each fold has 4 train subjects and 1 validation subject. 
All reported Dice scores (train and test) are averaged over these 5 folds. As the baseline UNet model, we adopt the multi-branch variant \cite{aslani2019multi} (MB-UNet) which utilizes the 4 modalities and all orthogonal planes of the 3D volumes to achieve competitive results for this task.


\begin{table*}
\begin{center}
\begin{tabular}{|l|c|c|c|c|}
\hline
Method &  Train Dice & Test Dice & Receptive Field & Params (M)\\
\hline\hline
MB-UNet $3\times 3$\cite{aslani2019multi} &  0.887 & 0.624 & 68 pixels & 2.1\\
MB-UNet $5\times 5$ &  0.893 & 0.605 & 128 pixels & 5.3\\
Dilated MB-UNet $3\times 3$ & 0.881 & 0.625 & 128 pixels & 2.1\\
HyperSeg \cite{nirkin2021hyperseg} & 0.864 & 0.64 & All pixels & 1.2 \\
Non-local MB-UNet $3\times 3$ \cite{nonlocalunet} & 0.905 & 0.637 & All pixels & 2.3\\
Hyper-MB-UNet $5\times 5$ (ours) & 0.82 & $\mathbf{0.655}$ & 128 pixels & 1.2\\
Non-local Hyper-MB-UNet $5\times 5$ (ours) & 0.834 & $0.652$ & All pixels & 1.4\\
\hline
Flat Dilated CNN\cite{yu2015dilation} & 0.854 & 0.616 & 89 pixels & 0.45\\
Flat Hyper-CNN (ours) & 0.805 & 0.649 & 89 pixels & 0.45\\
\hline
\end{tabular}
\end{center}
\caption{Train and Test Performance of different models in MS lesion segmentation task. Best test Dice score is \textbf{bold-faced}.}\label{tab3}
\end{table*}

\textbf{Results:}
Figure~\ref{brain} shows test Dice scores for the MB-UNet models with different numbers of parameters obtained by varying the number of channels and kernel size for standard convolution. 
Similar to previous experiments, the hyper-convolutions counterparts implement $5\times5 $ kernels that replace the $3\times 3$ regular convolutions.
As before,  hyper-convolutions consistently boost the test Dice score over a wide range of parameterizations, especially when the model size gets larger.   

Table~\ref{tab3} lists train and test results from the epoch with best validation loss, in addition to the receptive field size and number of learnable parameters, for baseline models and their hyper-convolution counterparts.
As above, we observe that hyper-convolutions boost test performance and shrink the gap between train and test loss compared to other strong baselines. 

Contrary to what we observed in liver lesion segmentation, the flat Hyper-CNN yields better results than the MB-UNet baseline and the non-local UNet, which have larger receptive fields. 
This difference is likely because MS lesions are relatively small and more local compared to liver lesions, and thus a larger receptive field is less beneficial for getting an accurate lesion segmentation (Figure~\ref{fig:exp}).  

\textbf{Hyperparameters:} 
Table~\ref{tab4} shows results for the 32-channel Hyper-MB-UNet with variable kernel sizes and $N_L$ values. 
Similar to liver lesion segmentation, we observe that the $5\times 5$ implicit kernel yields the best results.
We also note that, as before, increasing the capacity of the hyper-network does not always improve test performance, presumably due to overfitting. 
This underscores the importance of the regularization effect of hyper-convolutions.

\begin{table}[b]
\begin{center}
\begin{tabular}{|c|c|c|c|}
\hline
Size, $N_L$ & Test Dice & Recep. Field & Params (M)\\
\hline\hline
$3\times 3$, $2$ & 0.622 & 68 pixels & 0.73\\
$3\times 3$, $4$ & 0.625 & 68 pixels & 1.2\\
$3\times 3$, $8$ & 0.617 & 68 pixels & 2.2\\
\hline
$5\times 5$, $2$ & 0.648 & 128 pixels & 0.73\\
$5\times 5$, $4$ & 0.655 & 128 pixels & 1.2\\
$5\times 5$, $8$ & 0.651 & 128 pixels & 2.2\\
\hline
$7\times 7$, $2$ & 0.634 & 188 pixels & 0.73\\
$7\times 7$, $4$ & 0.644 & 188 pixels & 1.2\\
$7\times 7$, $8$ & 0.646 & 188 pixels & 2.2\\

\hline
\end{tabular}
\end{center}
\caption{Performance of Hyper-MB-UNet on MS Lesion data with different kernel sizes and hyper-network capacity.}\label{tab4}
\end{table}

\subsection{Segmentation with Noisy Images} \label{seg_noisy}
We also test our model on images with various types of added noise such as Gaussian and speckle. 
All the experiments are run on liver segmentation because there is more data to use. 
All models are trained with clean images, and all the hyper-parameters are kept consistent with previous experiments. 
During validation/inference, we add different types and level of noise to the input image to simulate noisy images acquired under different conditions. 
Figures ~\ref{noise} shows test Dice scores when we add salt and pepper, Gaussian, and speckle noise to the input images. 
As we already saw in the previous section, hyper-convolutions achieve better performance with clean validation images. 
When we add noise to the input images, regardless of the noise type, the performance gap between hyper-convolution and regular convolution increases with the amount of noise. 
Models with hyper-convolution are very robust against noise, even though they have only been trained noise-free images. 

%
%
%
%
%

\begin{figure}[tbh]
\begin{center}
\includegraphics[width=0.8\linewidth]{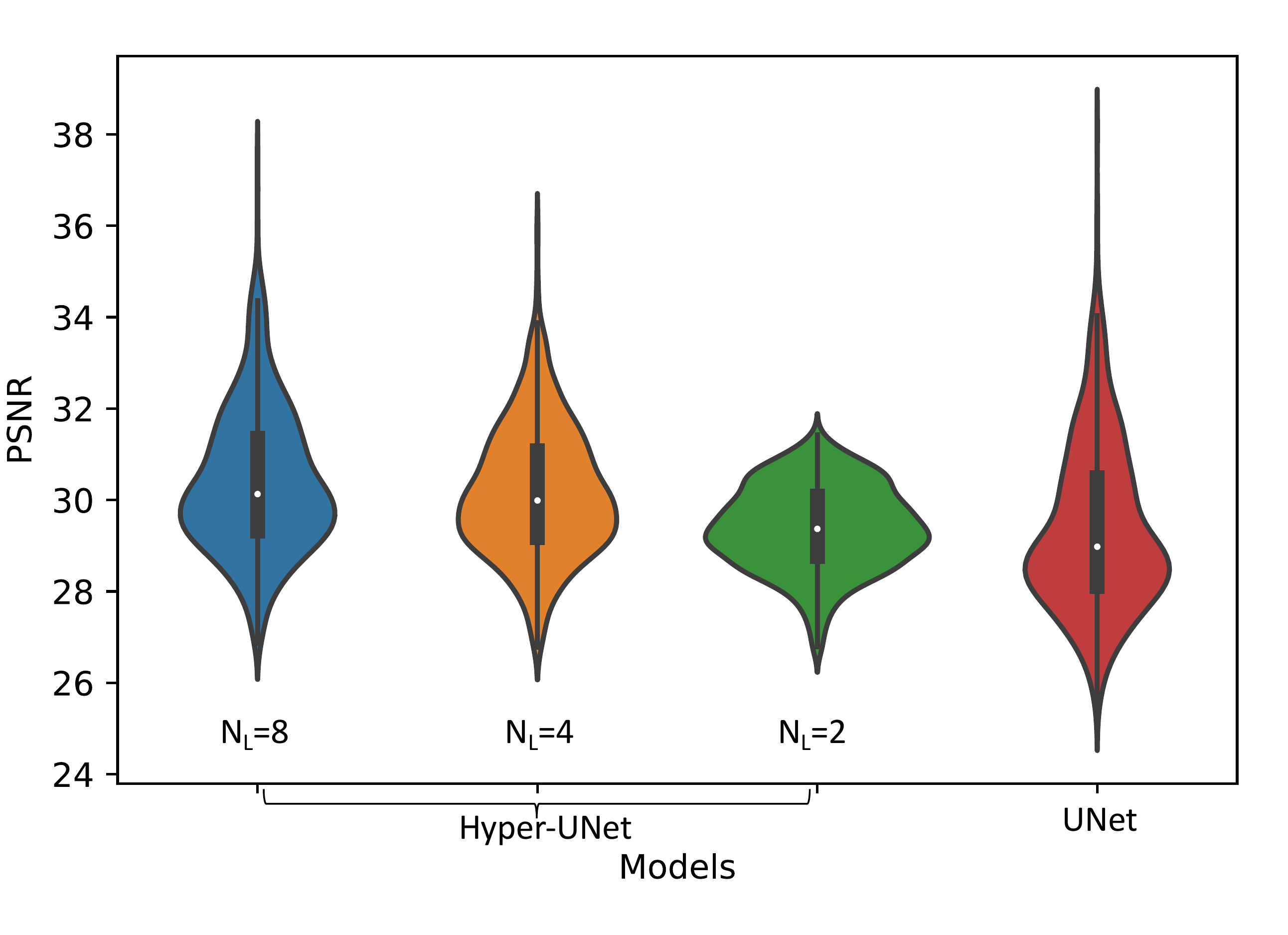}
\end{center}
   \caption{Violin plots for test PSNR of three Hyper-UNet models ($N_L=8,4,2$) and one UNet. }
\label{fig:violin}
\end{figure}

\begin{figure}[tbh]
\begin{center}
\includegraphics[width=0.7\linewidth]{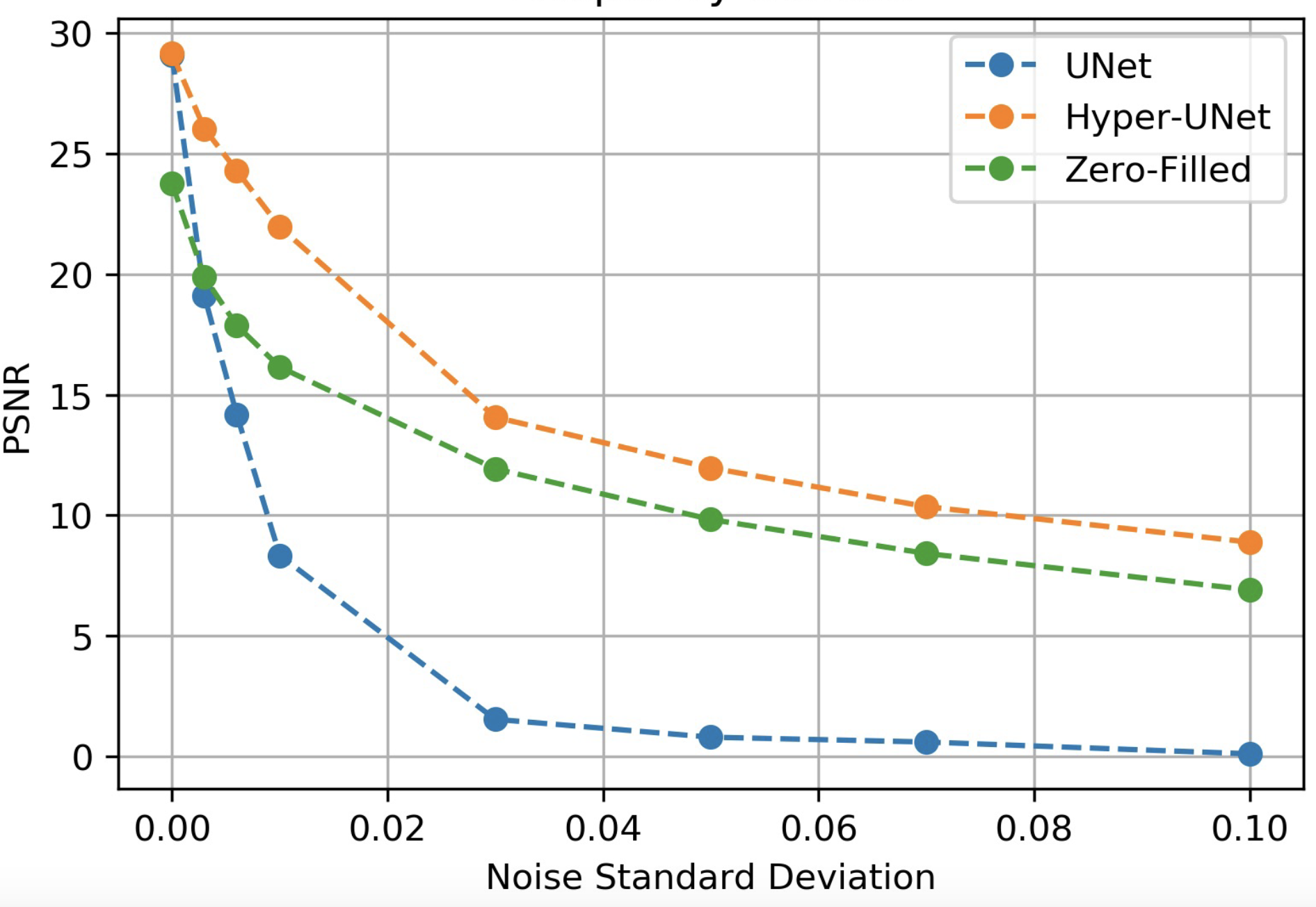}
\end{center}
   \caption{CS-MRI from noisy data: Test PSNR with different noise levels.}
\label{fig:psnr}
\end{figure}

\subsection{CS-MRI Reconstruction}
Hyper-convolution can be applied beyond the image segmentation problem. 
In this work, we also consider the compressed sensing MRI (CS-MRI) reconstruction task. 
CS-MRI seeks to perform $k$-space sampling at sub-Nyquist rates, allowing for accelerated data acquisition. Given an under-sampled $k$-space slice, the goal of reconstruction is to recover the corresponding fully-sampled image. 

In this work, we solve the reconstruction problem by inputting the zero-filled reconstruction\footnote{Defined as the inverse Fourier transform of the under-sampled k-space scan with zeros filled in for missing values.} into a neural network model, and training the parameters of the model to minimize the mean-squared-error against the corresponding fully-sampled image.
We use the same UNet architecture as above to compare hyper-convolution with regular convolution, although we stress that our method is applicable to any model that uses convolutions, including recently-popular model-based networks~\cite{aggarwal2019modl,shlezinger2020model,wang2020fullysampled,willard2021integrating}.

\textbf{Data:}
We use a dataset of T1-weighted axial brain MRI images with 3500 2D scans \cite{dalca2018anatomical}. 
The dataset is split into 2000, 500, and 1000 slices for training, validation, and testing.
We normalize image intensities to $[0,1]$ and re-sample to uniform pixel grids with size $256\times256$. 
The k-space data are generated by retrospectively under-sampling by 8-fold using a Poisson-disk variable-density sampling strategy \cite{geethanath2013compressed}.

\begin{figure*}[tb]
\begin{center}
\includegraphics[width=0.9\linewidth]{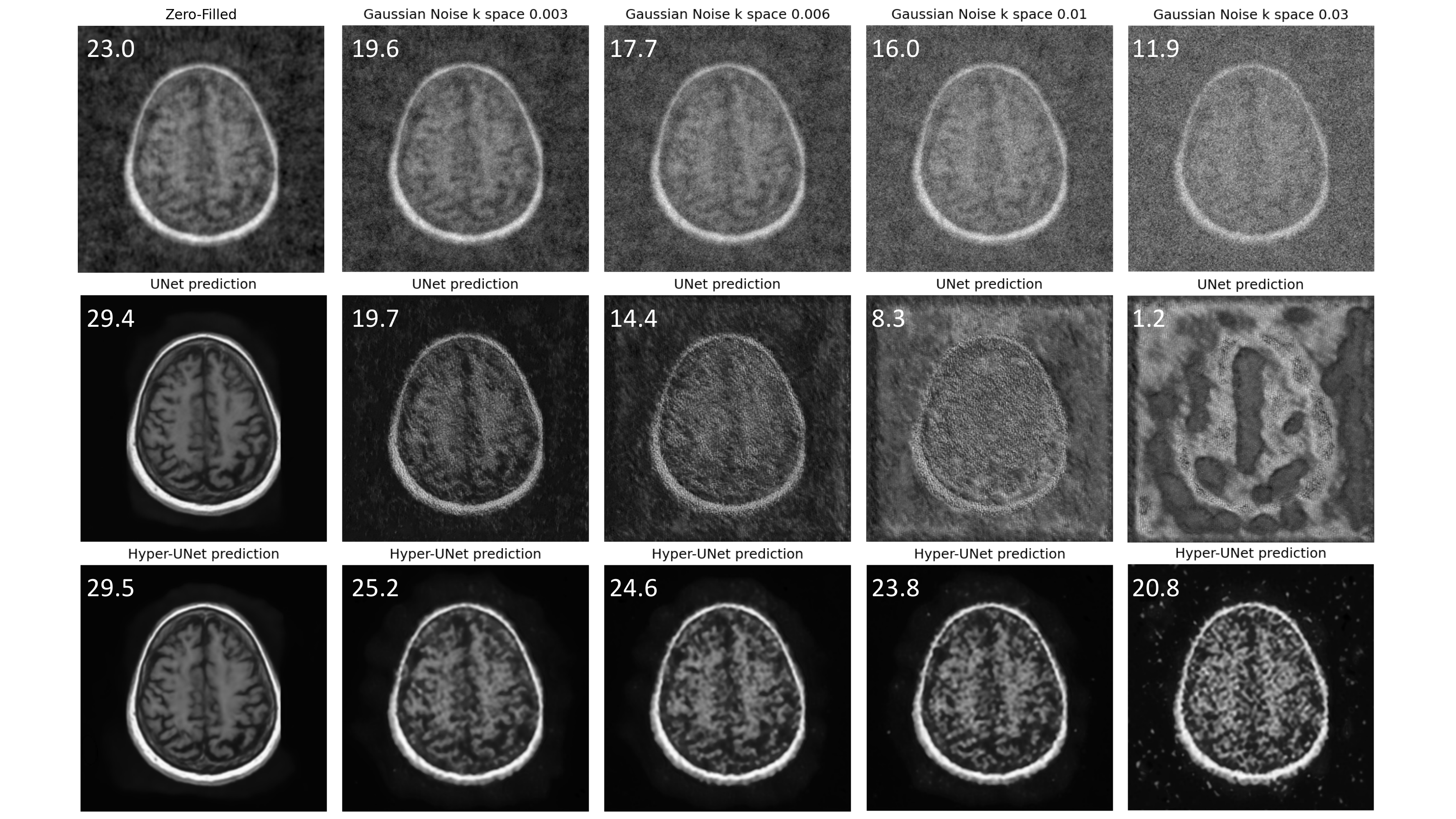}
\end{center}
   \caption{Test under-sampled images (top row), reconstructed images by UNet (middle row) and Hyper-UNet (bottom row) model at different Gaussian noise levels, added to frequency domain. Numbers are corresponding PSNR values. 
   }
\label{fig:recon_fre}
\end{figure*}

\textbf{Results:}
We first demonstrate the superior performance of hyper-convolution with fully-sampled test images. 
We train 3 Hyper-UNet models with $5\times5$ kernels and varying $N_L = 8,4,2$. The corresponding number of parameters for these three models are 2.2M, 1.2M, and 0.73M. Figure~\ref{fig:violin} shows violin plots for PSNR values. The average PSNR of these three models decrease from 30.4 to 30.16, 29.39, as the number of parameters decreases. 
The fourth model in Figure~\ref{fig:violin} is the baseline with regular $5 \times 5$ convolutions and 2.2M parameters. This baseline achieves an average PSNR of 29.36. We keep all other hyper-parameters such as number of channels the same for these four models to achieve a fair comparison. 
We observe that all hyper-convolution models outperform the baseline. Even the smallest hyper-convolution model obtains a similar PSNR to the baseline, with only 1/3 of the total trainable parameters.  
We also notice that hyper-convolution models have smaller variation in terms of the individual PSNR among all test images. 

We test the robustness of the reconstruction models (trained with clean data), by adding Gaussian noise (with different standard deviation) to the test data\footnote{Results shown correspond to adding the noise in k-space. Experiments with image domain noise exhibit same pattern of results and are excluded due to page restrictions.}
As reference, we show the PSNR for the zero-filled reconstructions, which are provided as input to the models. 
For a reconstruction network, the objective is to achieve an output PSNR value that is higher than the zero-filled (input) PSNR. 
We observe that both UNet and hyper-UNet obtain equally high PSNR values for noise-free input images. 
However, the performance of the UNet baseline quickly drops below the input PSNR as the degree of noise increases, underscoring the vulnerability of regular convolutions to even slight perturbations or corruption. 
On the other hand, the hyper-UNet yields output PSNR values that stay above the input PSNR, producing improved quality reconstructed images, despite the distribution shift introduced by the additive noise. 
A qualitative observation from Figure~\ref{fig:recon_fre} is that the hyper-convolution model can produce useful results, even when the baseline's output is completely wrong. 
We believe this type of robustness is critical for the real-world deployment of these models. 

%

\section{Discussion - Kernel Smoothness and Robustness}
\label{sec:discussion}


\begin{figure*}[tb]
\begin{center}
\includegraphics[width=0.8\linewidth]{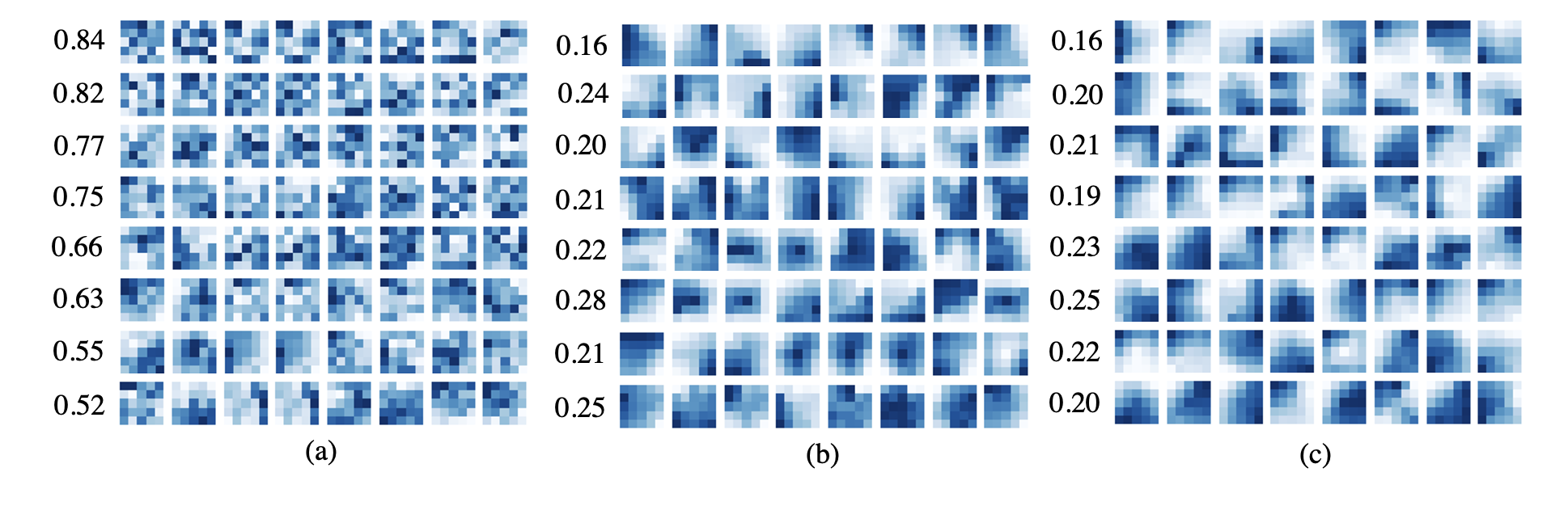}
\end{center}
   \caption{Visualization of $5\times 5$ convolutional kernels . Layerwise average Laplacians are listed for each of the networks: (a) UNet baseline, (b) Hyper-UNet ($N_L=8$), (c) Hyper-UNet ($N_L=24$). Each row corresponds to one network layer.}
\label{fig:kernel1}
\end{figure*}

\begin{figure}[tb]
\centering
\includegraphics[width=1\linewidth]{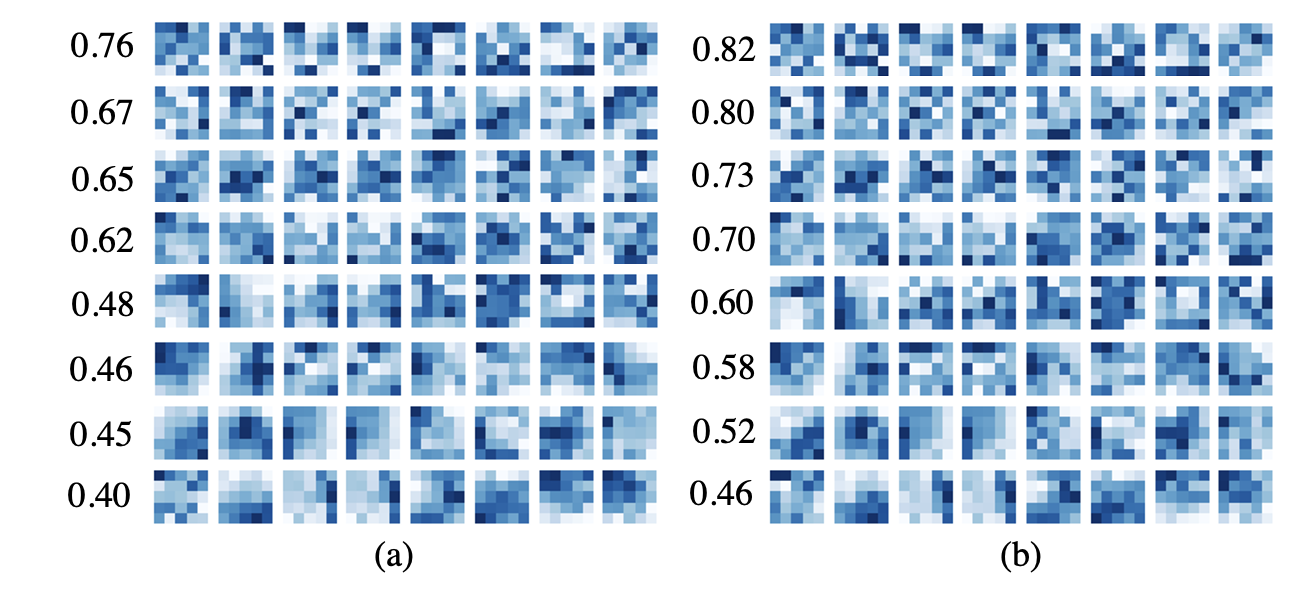}
\vspace{-0.5cm} 
   \caption{Implicit kernel approximations of the $5\times 5$ regular UNet kernels from Figure~\ref{fig:kernel1}(a): (a) $N_L=8$. (b) $N_L=24$. Each row corresponds to one network layer, with the average Laplacian value listed.}
\label{fig:kernel2}
\end{figure}

As mentioned above, we empirically observe that the learning dynamics of hyper-convolution yields smoother kernels, even if the number of parameters is not restrictive. 
To better characterize this, here, we perform an analysis of the kernels of the $5\times5$ UNet and Hyper-UNet models trained for the liver segmentation task. 
Figure~\ref{fig:kernel1} shows the learned kernels from standard convolutions (a) and hyper-convolutions (b,c). 
Each panel consists of 8 rows corresponding to sequential layers in the network.   
In each layer we show 8 randomly chosen kernels along with the smoothness of that layer. 
The smoothness of a kernel is quantified by calculating its average 2nd-order spatial derivative (Laplacian). Lower Laplacian values indicate smoother kernels.
We include results obtained from two hyper-convolution models with different capacities, and find out that the kernels learned in both the low capacity ($N_L=8$) and high capacity ($N_L=24$) Hyper-CNNs are significantly smoother than those learned in the standard UNet despite the high capacity hyper-convolution being equally expressive as the standard $5\times5$ convolution (as we show below).

Next, we conduct an analysis to test whether the smoothness of the implicit kernels is due to learning dynamics or limited expressiveness. 
We train a hyper-network to recapitulate each learned UNet kernel by minimizing L2 loss on kernel weight values.
While the high capacity hyper-convolution network yields a more accurate recapitulation, the low capacity model can also generate a good approximation, despite having only less than half of the parameters of the regular $5\times 5$ convolution kernel (Figure~\ref{fig:kernel2}). 

These results show that the limited capacity of a hyper-network cannot be the only explanation for the smooth implicit kernels, since there seems to be enough capacity to obtain non-smooth approximations of the regular learned kernels.
Instead, we conclude that the smooth implicit kernels (Figure~\ref{fig:kernel1}) are in part due to different learning dynamics.   

The benefits of smooth kernels have been widely studied in previous research. 
Feinman and Lake \cite{smooth} proposed to use a smooth kernel regularizer to encourage the kernel weights to be spatially correlated. They demonstrated that smooth kernels show better generalization performance. Recently, Wang \MakeLowercase{\textit{et al.}} \cite{wang2020high} also suggested that models with smooth convolutional kernels, especially in the earlier layers, tend to have better adversarial robustness. They argued that a smooth kernel can ignore the high-frequency component of an image which is usually invisible to the human eye, but can be disruptive for neural network predictions.
They trained regular convolution kernels and manually smoothed them out, and tested the performance with distorted images. 
In their experiments, models with smooth kernels show better robustness against adversarial attacks at the cost of lower performance with inputs without perturbation. 
In hyper-convolutions, the spatial smoothness in the learned kernels can be directly imposed by restricting the capacity of the hyper-network, yet we observe that even with high capacity the learned kernels exhibit spatial smoothness. 
We believe that this spatial smoothness explains the generalization and robustness of Hyper-CNNs.

As visualized in Figure~\ref{fig:kernel1}, we find that deeper layer kernels in the Hyper-CNN are less smooth than the first layer, as evidenced by higher Laplacian values. 
This can be particularly useful when there is a lot of high-frequency noise present in the input images.  
In our baseline CNN models, We do not observe such a pattern. Instead, the smoothness of kernels seems to increase with deeper layers. 

\section{Conclusion}
In this paper, we presented hyper-convolution, a novel building block that can be used with any convolutional neural network architecture to replace regular convolutional kernels. 
The hyper-convolution represents a convolutional kernel as an implicit function that maps grid coordinates to weight values. 
Hence, hyper-convolutions decouple the total number of learnable parameters in a kernel from its size and support, enabling the use of larger filters with greater receptive field without having too many learnable parameters. 
Furthermore, as our results demonstrate, the learned implicit kernels are often smoother than their regular counterparts, due in part to different learning dynamics, which can help combat overfitting and improve generalization and robustness. 
We believe hyper-convolutions will be an important building block for future neural network architectures, enabling researchers to further explore the trade-offs between capacity and generalization. 
{\small
\bibliographystyle{ieee_fullname}
\bibliography{egbib}
}

\end{document}